\documentclass[aps,pre,twocolumn,superscriptaddress,longbibliography,floatfix]{revtex4-2}
\usepackage{amsmath,amssymb,amsfonts,mathtools}
\usepackage{bm}
\usepackage{graphicx}
\usepackage{xcolor}
\usepackage{dcolumn}
\usepackage{hyperref}
\usepackage{xcolor}
\usepackage{physics}
\usepackage{bbm}

\begin{document}

\title{Probing Chaos and Criticality with Observational Entropy and Finite-Resolution Measurements}

\author{J. Bharathi Kannan}
\email{bharathikannan1130@gmail.com}
\affiliation{Department of Physics, Indian Institute of Science Education and Research, Pune 411008, India}

\author{Sreeram PG}
\email{sreerampg7@gmail.com}
\affiliation{Department of Physics, Indian Institute of Science Education and Research, Pune 411008, India}
\affiliation{School of Computing, MIT Art, Design and Technology University, Pune 412201, India}

\author{M.S. Santhanam}
\email{santh@iiserpune.ac.in}
\affiliation{Department of Physics, Indian Institute of Science Education and Research, Pune 411008, India}
\date{\today}

\begin{abstract}
Coarse-grained measurements offer a scalable alternative to full state tomography for characterizing complex quantum dynamics. We show that observational entropy (OE), an information-theoretic entropy defined directly from finite-resolution measurement outcomes, provides a unified and experimentally accessible framework for quantifying chaos and probing criticality. From probing the insulator–metal crossover in the Aubry–Andr\'e model to tracking the gradual destruction of Kolmogorov–Arnold–Moser tori in the Kicked Rotor, derivatives of OE provide an accurate and unified diagnostic of probing these transitions. In both cases, the critical points extracted from dynamical evolution and eigenstate analyses converge to the exact theoretical values once the observational resolution exceeds a finite threshold. In the chaotic limit, OE exhibits a linear behavior within the Ehrenfest time regime, and its slope defines an observable Lyapunov exponent. Using a Pretty Good Measurement correction to the Husimi phase-space distribution, this entropy-production rate quantitatively reproduces the classical Lyapunov exponent in both the standard and singular kicked rotors. Our results establish OE as a compact information-theoretic bridge between classical instability, quantum criticality, and realistic finite-resolution measurements.
\end{abstract}

\maketitle

\section{Introduction}
\label{sec:intro}

Understanding how classical chaotic behaviour emerges from unitary quantum mechanics remains a central problem in quantum physics \cite{gutzwiller1990chaos, haake1991quantum, chirikov1979universal, Stockmann1999}. In Hamiltonian systems, chaos is associated with exponential sensitivity to initial conditions, phase-space stretching, and, in near-integrable settings, resonance-driven destruction of invariant KAM tori \cite{arnold1963small,moser1962invariant}. In periodically kicked systems such as the kicked rotor (KR), this also leads to diffusive momentum transport. In contrast, unitary quantum evolution is linear and norm-preserving, so classical trajectory divergence is not directly realised in Hilbert-space state distance. As a result, semiclassical trajectory-based signatures of chaos persist only up to the Ehrenfest time, beyond which quantum interference effects dominate \cite{berry1979evolution, zaslavsky1981stochasticity}. Several established spectral and dynamical probes capture complementary signatures of quantum chaos. Spectral statistics distinguish chaotic and integrable systems through level-spacing distributions  \cite{bohigas1984characterization, berry1977level, berry1984semiclassical}. Dynamical measures such as the Loschmidt echo, quantum fidelity, and out-of-time-ordered correlators probe sensitivity to perturbations, operator growth, and scrambling, with early-time behaviour often reflecting Lyapunov-like signatures \cite{peres1984stability, jalabert2001environment, gorin2006dynamics, cucchietti2003decoherence, larkin1969quasiclassical, maldacena2016bound, rozenbaum2017lyapunov, shenker2014black, GarciaMata2023OTOCs}. While powerful, these methods often require extensive spectral information, precise control protocols, or access to nonlocal correlators.\\

A complementary approach is to characterise dynamics using entropy measures that explicitly account for finite measurement resolution. Coarse-graining plays a central role in statistical irreversibility, consistent-histories formulations, and black-hole information theory \cite{gell1993classical, griffiths1984consistent, almheiri2020page, penington2020entanglement}. Within this framework, observational entropy (OE) provides a natural resolution-dependent measure of accessible information \cite{vsafranek2019quantum,vsafranek2019quantum2,vsafranek2021brief}. OE interpolates between or reproduces von Neumann, Gibbs, and Boltzmann entropies in appropriate limits, and approaches the Kolmogorov–Sinai entropy rate in the appropriate classical limit \cite{vsafranek2019quantum, vsafranek2019quantum2, latora1999kolmogorov}. It has also been used to study thermalisation and ergodicity breaking in many-body systems \cite{pg2023witnessing,buscemi2023observational,leblond2019entanglement,pappalardi2023quantum}. Despite these advances, a systematic understanding of how much physically relevant information can be extracted from OE under finite measurement resolution remains incomplete. \\

A central question motivating this work is the practical utility of OE: can an information-theoretic quantity defined at finite measurement resolution be used to extract experimentally relevant signatures of chaos, ergodicity breaking, and criticality? We show that OE is  useful as a unified, operational diagnostic that quantitatively captures phenomena ranging from the gradual destruction of Kolmogorov-Arnold-Moser (KAM) tori in the Kicked rotor (KR) to the self-dual transition in the Aubry-Andr\'e (AA) model. Our results reveal that the order of the derivative of OE required to detect a transition encodes the underlying mechanism of ergodicity breaking: the smooth breakdown of KAM barriers produces a singularity in the second derivative of OE, whereas the AA localization transition appears directly in the first derivative. Finite-size scaling of these entropy derivatives yields the exact critical points and the correct critical exponent, while also exposing an intrinsic resolution bound beyond which the extracted values become insensitive to further refinement of the coarse-graining. \\

We further show that, within the Ehrenfest time scale, OE grows linearly in time in both the standard kicked rotor (KR), where chaos emerges through the progressive destruction of KAM tori, and the singular kicked rotor (SKR), a paradigmatic non-KAM system with abrupt transition to chaos. In both cases, the growth rate quantitatively reproduces that corresponding to the classical Lyapunov exponent. A Husimi phase-space coarse-graining, based on the positive-operator-valued measure (POVM) of coherent states, yields a significantly more accurate estimate of this rate than momentum-space binning. Since Husimi distributions can be reconstructed experimentally via ancilla-assisted measurements of coherent-state overlaps, the corresponding Husimi-based OE is directly accessible in quantum simulators. Taken together, these results establish OE as a unified, resolution-aware framework for extracting quantitative signatures of chaos, localization, and criticality from realistic coarse-grained measurements in platforms such as cold-atom lattices \cite{moore1995atom,chabe2008experimental,lopez2012experimental,maurya2022interplay,sunil2026localization}, quasiperiodic ultracold systems \cite{roati2008anderson,luschen2018single}, and Floquet superconducting circuits \cite{roushan2017spectroscopic, mi2021information}.\\

The remainder of this paper is organised as follows. In Sec.~\ref{sec:formalism}, we introduce the formalism of standard OE for orthogonal coarse-grainings, applicable to momentum space, position space, and more general orthogonal subspaces, and we define the models studied in this work. In Sec.~\ref{sec:criticality}, we show how OE can be used to probe dynamical and localisation transitions, using the quantum kicked rotor and Aubry-Andr\'e models as representative examples. In Sec.~\ref{sec:lyapunov}, we extend the framework to phase-space coarse-graining and show how Lyapunov-exponent signatures can be extracted in Husimi space using the Pretty Good Measurement (PGM) formalism, comparing these results with momentum-space estimates along. Finally, in Sec.~\ref{sec:conclusions}, we summarise the main results and discuss future directions.

\section{Formalism and Models}
\label{sec:formalism}

\paragraph{Observational Entropy.}
Consider a quantum system with Hilbert space $\mathcal{H}$ of dimension $d$. A coarse-graining $\chi = \{\Pi_i\}$
is a complete set of mutually orthogonal projectors satisfying $\Pi_i \Pi_j = \delta_{ij}\Pi_i, \sum_i \Pi_i = \mathbbm{1}.$ Each projector $\Pi_i$ defines a macrostate of dimension $V_i = \Tr(\Pi_i).$ For a state $\rho$, the probability of observing macrostate $i$ is $p_i = \Tr(\Pi_i \rho).$
The observational entropy associated with the coarse-graining $\chi$ is \cite{vsafranek2019quantum}
\begin{equation}
S_{\chi}(\rho)
=
-\sum_i p_i \ln\!\left(\frac{p_i}{V_i}\right)
=
-\sum_i p_i \ln p_i
+
\sum_i p_i \ln V_i .
\label{eq:OE_def}
\end{equation}
The first term is the Shannon entropy of the macrostate distribution, while the second term accounts for the unresolved microscopic degeneracy within each macrostate. Thus, $S_{\chi}(\rho)$ quantifies the uncertainty seen by an observer with finite resolution. Equation~\eqref{eq:OE_def} interpolates between two limiting cases. If all projectors are rank one ($V_i=1$), observational entropy reduces to the Shannon entropy of the measurement outcomes. If the coarse-graining consists of a single projector $\Pi=\mathbbm{1}$, then $S_{\chi}(\rho)=\ln d$, corresponding to the microcanonical Boltzmann entropy.\\

Observational entropy is monotonic under further coarse-graining, reflecting the fact that reducing measurement resolution can only increase uncertainty. Moreover, although the von Neumann entropy remains constant under unitary evolution, $S_\chi(\rho)$ can increase as the state spreads over multiple macrostates, providing an operational formulation of entropy growth without assuming ergodicity \cite{vsafranek2019quantum2}. In this work, we partition the Hilbert space of dimension $N$ into $n_{\mathrm{sub}} = \frac{N}{\chi}$ equal macrostates, each of dimension $V_i = \chi$. The integer $\chi$ therefore sets the observational resolution: small $\chi$ corresponds to fine-grained measurements sensitive to quantum interference, while large $\chi$ probes only coarse macroscopic structure. For phase-space coarse-graining, the same construction is applied to the PGM-corrected Husimi distribution, where macrostates correspond to cells in the $(q,p)$ plane. More details on this will be discussed in section \ref{sec:lyapunov}. In the first part of this work, we use Observational Entropy to probe ergodicity-breaking transitions in two paradigmatic models. The quantum kicked rotor captures the gradual onset of global chaos through the destruction of KAM tori, while the Aubry--Andr\'e model exhibits a self-dual transition from extended to exponentially localized eigenstates.\\

\paragraph{Standard Kicked Rotor (KR).} The kicked rotor \cite{casati2005stochastic,izrailev1990simple,santhanam2022quantum} describes a particle on a ring subjected to periodic $\delta$-kicks:
\begin{equation}
\hat{H}(t)
=
\frac{\hat{p}^2}{2}
+
K \cos \hat{x}
\sum_{n\in\mathbb{Z}}
\delta(t-n),
\label{eq:QKR_H}
\end{equation}
where $K$ is the kick strength. In the classical limit, Eq.~\eqref{eq:QKR_H} reduces to the Chirikov standard map \cite{chirikov1979universal}. For $K < K_c \simeq 0.9716$, invariant KAM tori inhibit global transport, while for $K > K_c$ resonance overlap leads to widespread chaotic diffusion \cite{greene1979method}. Quantum evolution initially follows the classical dynamics up to the Ehrenfest time, $t_E \sim \frac{\ln(1/\hbar)}{2\lambda}$ where $\lambda$ is the classical Lyapunov exponent. For $K \gg 1$, the momentum variance grows diffusively, $\langle p^2(t) \rangle \approx D t$ and 
$D \simeq \frac{K^2}{2}.$ This growth persists until the break time $t^\ast \sim \ell$ where $\ell \propto \frac{D}{\hbar^2} \sim
\frac{K^2}{\hbar^2}$ is the localization length in momentum space. Beyond $t^\ast$, destructive interference suppresses further diffusion, producing \emph{dynamical localization} \cite{fishman1982chaos,grempel1984quantum}. In this regime, Floquet eigenstates are exponentially localized, $|\psi_n|
\propto e^{-|n-n_0|/\ell}$ and the kinetic energy saturates. This phenomenon is formally equivalent to Anderson localization in one dimension and has been observed experimentally in cold-atom realizations of the kicked rotor \cite{moore1995atom,chabe2008experimental,santhanam2022quantum}.\\

\paragraph{Singular Kicked Rotor (SKR).} A non-KAM generalization of the QKR is obtained by replacing the smooth cosine kick with a nonanalytic power-law potential \cite{garcia2005anderson},
\begin{equation}
\hat{H}(t)
=
\frac{\hat{p}^{2}}{2}
+
\epsilon |\hat{x}+b|^{\alpha}
\sum_{n\in\mathbb{Z}}\delta(t-n),
\qquad
\alpha \in [-1,1].
\label{eq:SKR_H}
\end{equation}
In the numerics, we use the equivalent periodic regularization
\begin{equation}
V(x)
=
\epsilon
\left(
\sqrt{2(1-\cos x)+\delta^2}
+
|b|
\right)^{\alpha},
\label{eq:SKR_periodic}
\end{equation}
which reduces to $V(x)\sim \epsilon (|x|+|b|)^\alpha$ near $x=0$ and is well defined on the circle. Here $\delta \ll 1$ is a numerical cutoff.
The exponent $\alpha$ controls the degree of nonanalyticity, while $b$ tunes the effective strength of the singularity. For $b=0$, the singularity is strongest; increasing $|b|$ progressively regularizes the potential and suppresses the associated instability. In this work, $b$ is used as the primary control parameter.\\

Unlike the standard KR, the singular kicked rotor exhibits anomalous transport for $-1/2 < \alpha < 1/2$, characterized by heavy-tailed momentum distributions and L\'evy flights \cite{zaslavsky2002chaos}. In the quantum regime, Floquet eigenstates develop power-law momentum tails,
$|\psi_n|^2 \propto |n-n_0|^{-\eta(\alpha)}$,
reflecting the long-range couplings induced by the singular potential \cite{garcia2005anderson}; the case $\alpha=0$ is critical and supports multifractal eigenstates. Because such dynamics is not fully captured by low-order moments such as $\langle p^2(t)\rangle$, the SKR provides a stringent test of observational entropy. In Sec.~\ref{sec:lyapunov}, we compare the growth rates of momentum-space and PGM-corrected Husimi observational entropies with the classical Lyapunov exponent as a function of $b$, quantifying how finite-resolution measurements capture chaos in systems even with singular dynamics.\\

\paragraph{Aubry-Andr\'e Model.} The Aubry-Andr\'e (AA) model \cite{aubry1980analyticity,harper1955single} describes a one-dimensional tight-binding lattice with a quasiperiodic onsite potential,
\begin{equation}
\hat{H}
=
-J\sum_{j}
\left(
|j\rangle\langle j+1|
+
\mathrm{H.c.}
\right)
+
\lambda
\sum_{j}
\cos(2\pi\beta j+\phi)
|j\rangle\langle j|,
\label{eq:AA}
\end{equation}
where $J$ is the nearest-neighbor hopping amplitude, $\lambda$ is the modulation strength, $\phi$ is an arbitrary phase, and $\beta$ is irrational (typically $\beta = (\sqrt{5}-1)/2$), ensuring that the potential is incommensurate with the lattice. The model is exactly self-dual under Fourier transformation, which fixes the localization transition at $\lambda_c = 2J$. All eigenstates are extended for $\lambda < 2J$, critical at $\lambda = 2J$, and exponentially localized for $\lambda > 2J$ with localization length $\xi =\frac{1}{\ln\!\left(\frac{\lambda}{2J}\right)}$ and near the transition,
$\xi \sim |\lambda-2J|^{-1}.$ Unlike one-dimensional systems with uncorrelated disorder, where all states are localized for arbitrarily weak randomness, the AA model exhibits a sharp global localization transition without a mobility edge. Its simplicity, exact solvability, and direct realization in ultracold atoms and photonic lattices make it a paradigmatic model for studying quasiperiodic localization and criticality \cite{roati2008anderson,luschen2018single,kraus2012topological}.\\

\section{Criticality and Resolution in Observational Entropy}
\label{sec:criticality}

In this section, we show that observational entropy provides a unified probe of ergodicity-breaking transitions in both dynamical evolution and eigenstate (or Floquet-state) ensembles. Beyond locating the transition points, the structure of the entropy derivatives distinguishes different mechanisms of ergodicity breaking. In the kicked rotor, chaos emerges progressively through the hierarchical destruction of invariant KAM tori, leading to a smooth entropy crossover accompanied by a pronounced peak in the second logarithmic derivative. In contrast, the Aubry-Andr\'e model exhibits a sharper localization transition at the self-dual point, where the entropy develops a kink and the corresponding first derivative becomes singular. These distinct derivative signatures consistently appear in both the long-time dynamical entropy and the entropy computed from eigenstate or Floquet states.\\

\begin{figure}[t]
    \centering
    \includegraphics[width=\linewidth]{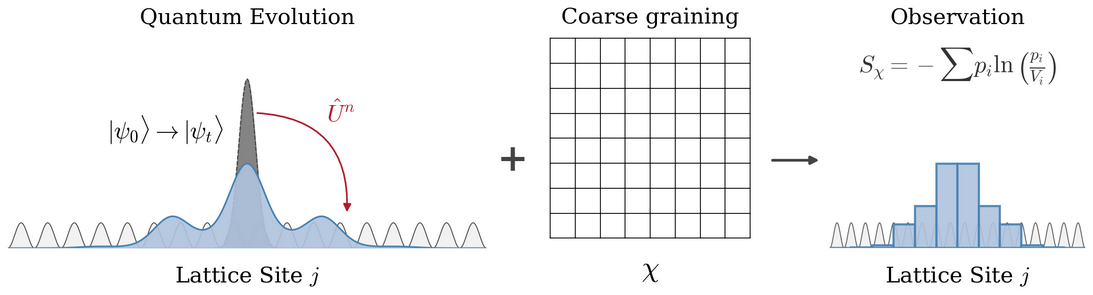}
    \includegraphics[width=\linewidth]{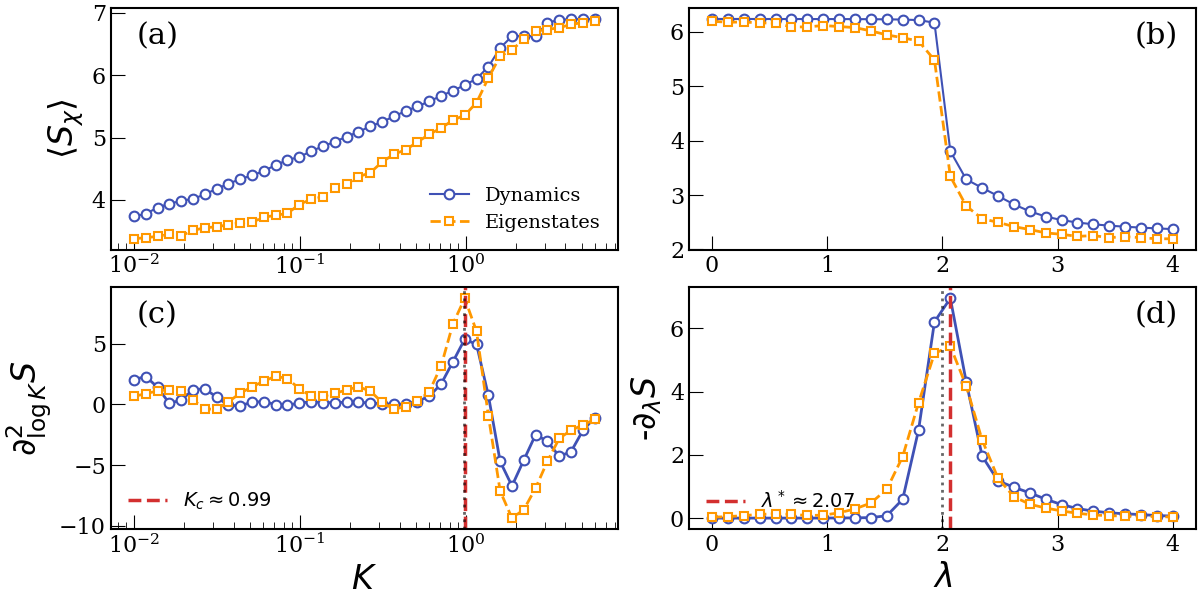}
    \caption{
    Top: Schematic of measuring observational entropy for a time-evolved state with coarse-graining length $\chi$. Bottom: (a) quasi-momentum averaged observational entropy $\langle S_\chi \rangle$ of the QKR versus kick strength $K$. (b) Entropy acceleration $\mathcal{A}_\chi(K)$, which peaks near the classical KAM threshold. (c) $\langle S_\chi \rangle$ of the Aubry-Andr\'e model versus $\lambda$. (d) Entropy susceptibility $\xi_\chi(\lambda)$, which dips near the self-dual localization transition. In all panels, results from long-time dynamics and eigenstate averages are in close agreement.}
    \label{fig:criticality}
\end{figure}

The top panel of Fig.~\ref{fig:criticality} illustrates the protocol used throughout this section to compute observational entropy in a dynamically accessible setting. Starting from an initially localized wave packet, the state is evolved under the corresponding unitary dynamics, after which the resulting probability distribution is coarse-grained into macrostates of resolution $\chi$. The observational entropy, $\langle S_\chi \rangle$, is then obtained by averaging over time-evolved states or ensembles of eigenstates/Floquet states. In the quantum kicked rotor, the ensemble average is taken over quasi-momentum sectors, while in the Aubry-Andr\'e model it is taken over the phase offset of the quasiperiodic potential. \\

In the KR, $\langle S_\chi \rangle$ increases smoothly with $K$, with a noticeable change in curvature near the onset of global chaos, as shown in Fig.~\ref{fig:criticality}(a). In the Aubry-Andr\'e model Fig.~\ref{fig:criticality}(b), $\langle S_\chi \rangle$ remains large in the extended phase and decreases sharply near the self-dual point as the eigenstates localize and occupy fewer distinguishable macrostates. In both models, the long-time dynamical and eigenstate results closely coincide, demonstrating that time evolution and stationary-state structure encode the same coarse-grained information. To locate the transition points, we analyze derivatives of $\langle S_\chi \rangle$ with respect to the relevant control parameter. For the KR, the clearest signature appears in the entropy curvature,
$\frac{\partial^2 \langle S_\chi \rangle}{\partial (\log K)^2}$ shown in Fig.~\ref{fig:criticality}(c). Its peak at $K^\ast \approx 0.99$ identifies the onset of global chaos and approaches very close to the classical KAM tori breaking point $K_c \simeq 0.9716$. For the Aubry-Andr\'e model, the relevant quantity is the entropy susceptibility, $-\frac{\partial \langle S_\chi \rangle}{\partial \lambda}$,
displayed in Fig.~\ref{fig:criticality}(d), whose peak around the self-dual transition at $\lambda_c = 2$. 
These results are obtained for a finite system size $N=2^{10}$ and finite coarse-graining resolution $\chi=2^{6}$. Despite these limitations, observational entropy yields quantitatively accurate estimates of the critical points, demonstrating its robustness as a finite-resolution diagnostic of ergodicity-breaking transitions. These results further show that the derivative order at which OE becomes singular reflects the nature of the transition itself: a second-derivative singularity for the gradual destruction of transport barriers in the KR, and a first-derivative singularity for the abrupt global localization transition in the Aubry-Andr\'e model.\\

\begin{figure}[t]
    \centering
    \includegraphics[width=\linewidth]{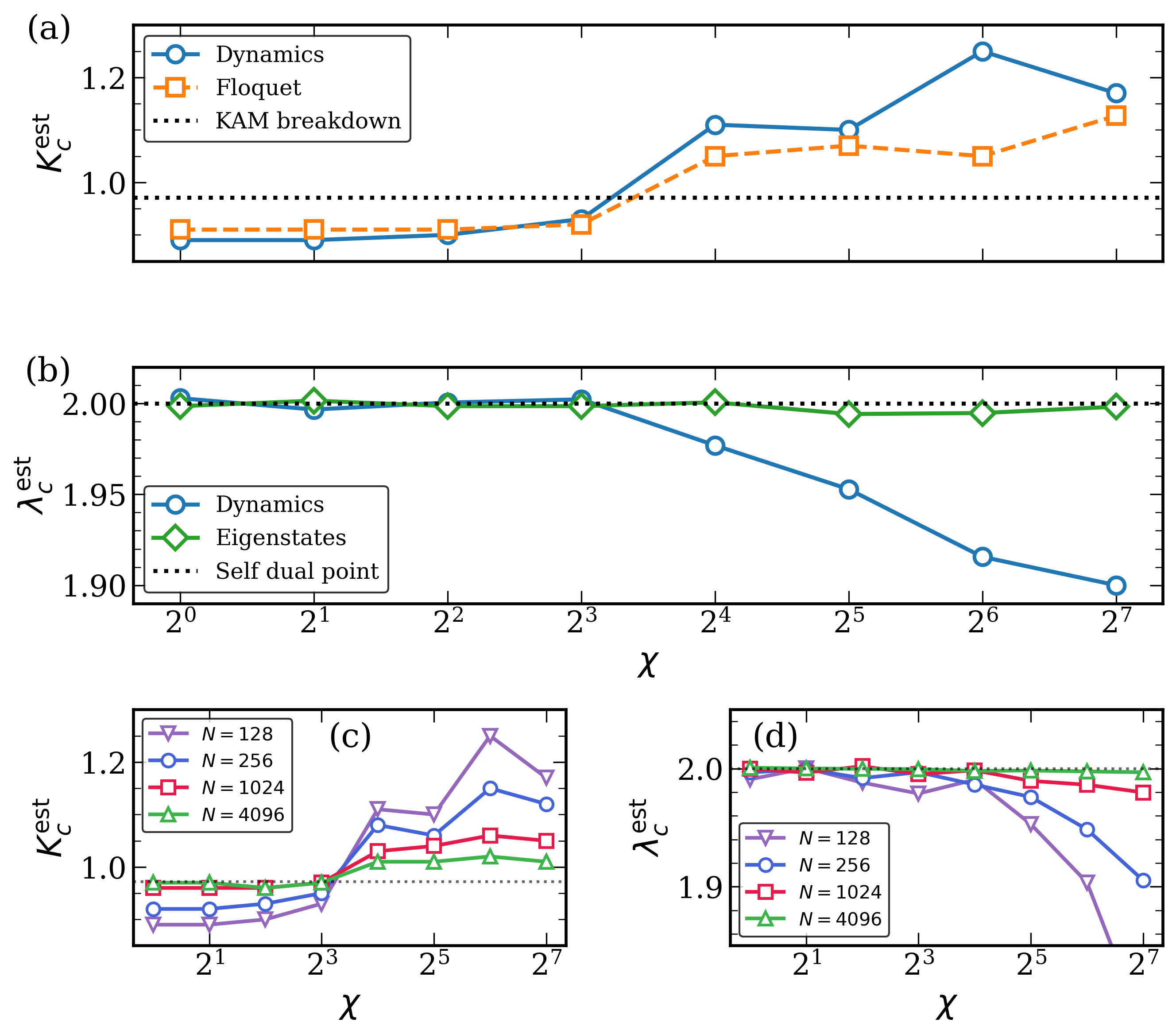}
    \caption{Critical-point estimates obtained from derivatives of the observational entropy as a function of the coarse-graining length $\chi$. (a) $K_c^{\mathrm{est}}$ for the quantum kicked rotor from the peak of the entropy curvature $\partial_K^2 \langle S_\chi \rangle$ (b) Estimated $\lambda_c$ for the Aubry-Andr\'e model from the peak of the entropy susceptibility $-\partial_\lambda \langle S_\chi \rangle$, comparing dynamical and eigenstate averages. Dotted lines denote the exact critical points. (c,d) Finite-size scaling of the corresponding estimates for different Hilbert-space sizes $N$. For experimentally relevant sizes ($N=128$), OE already yields accurate critical points, which converge rapidly to the exact values as $N$ increases.}
    \label{fig:transition_detection}
\end{figure}

We now examine how the extracted critical points depend on the coarse-graining length $\chi$. For each $\chi$, we locate the dominant peak of the corresponding derivative and denote the resulting estimates by $K_c^{\mathrm{est}}$ and $\lambda_c^{\mathrm{est}}$. Figure~\ref{fig:transition_detection} shows that both models exhibit a broad range of coarse-graining scales over which the OE-based estimates remain close to the exact transition points. For the experimentally relevant system size $N=2^7$, the critical points obtained from long-time dynamics and from Floquet/eigenstate ensembles already agree well with the theoretical values, $K_c \simeq 0.9716$ and $\lambda_c = 2$, demonstrating that OE retains the essential critical information even under finite observational resolution. Noticeable deviations appear only when the coarse graining becomes too large, at which point averaging over broad macrostates smooths the singular features in $\langle S_\chi \rangle$ and shifts the derivative peaks.\\

The finite-size analysis in Fig.~\ref{fig:transition_detection}(c,d) shows that these estimates converge systematically toward the exact critical points as $N$ increases, while the range of reliable coarse-graining scales broadens. Together, these results clearly show that derivative-based observational entropy provides a robust and experimentally realistic method for locating these transitions. The derivative order at which the singularity appears distinguishes the nature of the transition, whereas the $\chi$ dependence of the extracted critical point quantifies the measurement resolution required to resolve it accurately.

\section{Lyapunov Instability from Observational Entropy}
\label{sec:lyapunov}

The maximal Lyapunov exponent, $\lambda_{\mathrm{cl}}$, quantifies the fastest exponential separation of nearby classical trajectories, $\|\delta z(t)\| \sim \|\delta z(0)\| e^{\lambda_{\mathrm{cl}} t}$,
where $z=(q,p)$ denotes a phase-space point, and serves as a standard diagnostic of classical chaos. For sufficiently smooth ergodic systems, Pesin's theorem identifies the Kolmogorov-Sinai entropy rate with the sum of positive Lyapunov exponents, thereby linking dynamical instability to information production \cite{pesin1977characteristic,latora1999kolmogorov,ott2002chaos}. A central question is whether the classical Lyapunov exponent can be recovered directly from quantum measurements. Standard quantum probes of chaos capture complementary signatures of dynamical instability, but extracting Lyapunov exponents from them is often experimentally demanding. Out-of-time-ordered correlators can exhibit early-time exponential growth in suitable semiclassical regimes, allowing the extraction of an effective quantum Lyapunov exponent, but their measurement typically requires time-reversal protocols and access to nonlocal operators \cite{rozenbaum2017lyapunov,larkin1969quasiclassical,maldacena2016bound}. The Loschmidt echo can display a decay rate governed by the classical Lyapunov exponent in appropriate semiclassical regimes, but it relies on controlled forward and backward evolution under slightly perturbed Hamiltonians \cite{peres1984stability,jalabert2001environment}.\\

In this section, we show that within the Ehrenfest window, OE grows linearly in time with a rate that closely matches the classical Lyapunov exponent, providing a direct and experimentally accessible probe of dynamical instability. The mechanism is straightforward to understand: an initially localized minimum-uncertainty wave packet occupies a phase-space area of order $\hbar$. Under chaotic dynamics, the initially localized coherent wave packet, with extent $\delta \ell(0)\sim \sqrt{\hbar}$ along the unstable direction, is stretched exponentially as $\delta \ell(t)\sim \delta \ell(0)e^{\lambda_{\mathrm{cl}} t}$ while preserving its fine-grained phase-space area. A coarse-grained measurement with resolution $\chi$ cannot resolve structures smaller than a single phase-space cell. As the wave packet stretches along the unstable direction, it overlaps an increasing number of coarse-grained cells, $N_\chi(t)\sim \frac{\delta \ell(t)}{\chi}\propto e^{\lambda_{\mathrm{cl}} t}$. For the KR, where the Kolmogorov-Sinai entropy rate equals the maximal Lyapunov exponent \cite{latora1999kolmogorov}, this exponential growth directly produces a linear increase of observational entropy. As a result,
\begin{equation}
S_\chi(t)
\simeq
\ln N_\chi(t)
\simeq
S_0+\lambda_{\mathrm{cl}} t,
\qquad
t_{\mathrm{mix}}<t<t_E,
\label{eq:oe_linear_growth}
\end{equation}
where $t_{\mathrm{mix}}$ is a short transient and $t_E \sim \frac{1}{\lambda_{\mathrm{cl}}} \ln\!\left( \frac{\sigma_{\mathrm{cl}}}{\sqrt{\hbar}} \right)$ is the Ehrenfest time, with $\sigma_{\mathrm{cl}}$ denoting the characteristic classical scale over which the dynamics remains locally linear.\\

This motivates the observable Lyapunov exponent,
\begin{equation}
\lambda_{\mathrm{OE}}
=
\frac{dS_\chi}{dt},
\qquad
t_{\mathrm{mix}}<t<t_E,
\label{eq:loe_def}
\end{equation}
obtained from the slope of the linear growth regime. Because Lyapunov instability is inherently a phase-space phenomenon, accurately recovering $\lambda_{\mathrm{cl}}$ requires a coarse-graining that resolves both position and momentum. A single-basis partition, such as momentum-space binning, captures only a projection of the spreading. We therefore use the Husimi representation, which resolves the state into minimum-uncertainty coherent states localized in $(q,p)$.\\

Because coherent states form an overcomplete basis, the corresponding Husimi weights are neither independent nor properly normalized measurement probabilities. To construct an operational probability distribution, we apply the Pretty Good Measurement (PGM) \cite{fuchs2002cryptographic, hausladen1994pretty}, which converts the coherent-state ensemble into a normalized positive-operator-valued measure (POVM) \cite{nielsen2010quantum}. The resulting PGM-corrected Husimi distribution defines a bona fide phase-space probability distribution from which observational entropy can be computed.
We now compare two coarse-graining strategies for the KR: direct binning of the momentum distribution and phase-space coarse-graining of the PGM-corrected Husimi distribution. Momentum-space binning captures only a one-dimensional projection of the dynamics, whereas the PGM-Husimi construction resolves the full phase-space spreading of the wave packet. Comparing these two approaches reveals how much information about the classical Lyapunov exponent is retained in a single-observable measurement relative to a genuine phase-space measurement.

% \subsubsection{Momentum vs Phase-Space Coarse-Graining.}\\

In cold-atom realizations of the KR, the experimentally accessible observable is typically the momentum distribution, obtained from time-of-flight measurements \cite{santhanam2022quantum, maurya2022interplay}. It is therefore natural to ask whether the classical Lyapunov exponent can be extracted directly from momentum-space observational entropy, or whether a genuine phase-space coarse-graining is required. As we show below, momentum-space measurements capture the qualitative growth of entropy but systematically underestimate the instability rate, whereas the Husimi-based phase-space construction accurately reproduces the classical Lyapunov exponent. To extract the Lyapunov exponent from observational entropy, we compare two coarse-graining strategies. In momentum-space coarse-graining, the momentum basis is partitioned into bins of width $\chi$, and the resulting entropy growth rate defines
\begin{equation}
\lambda_{\mathrm{OE}}^{(\mathrm{mom})} = \frac{d S_\chi^{(p)}}{dt}.
\label{eq:lambda_mom}
\end{equation}
As this construction integrates over position, it captures only a one-dimensional projection of the phase-space dynamics.

To resolve the full stretching of the wave packet, we employ a phase-space coarse-graining based on minimum-uncertainty coherent states $\ket{\alpha}\equiv\ket{q_0,p_0}$ \cite{glauber1963coherent},
\begin{equation}
\langle q \mid \alpha \rangle
=
\left(\frac{1}{2\pi\sigma^2}\right)^{1/4}
\exp\!\left[-\frac{(q-q_0)^2}{4\sigma^2}\right]
\exp\!\left[\frac{i p_0(q-q_0)}{\hbar}\right],
\label{eq:coherent_state}
\end{equation}
with $\sigma=\sqrt{\hbar/2}$, so that $\Delta q\,\Delta p=\hbar/2$. The squared coherent-state overlaps $\langle \alpha_{kl}|\psi\rangle$, evaluated efficiently using the short-time Fourier transform (STFT), define the Husimi distribution $Q(q_k,p_l)=|\langle \alpha_{kl}|\psi\rangle|^2$
which provides a positive phase-space representation of the quantum state \cite{husimi1940some}.

Because coherent states are nonorthogonal, the raw Husimi weights are positive but do not constitute a normalized set of independent measurement probabilities. We therefore construct the Pretty Good Measurement (PGM) \cite{hausladen1994pretty, fuchs2002cryptographic} from the frame operator
\begin{equation}
G = \sum_{k,l} \ket{\alpha_{kl}}\!\bra{\alpha_{kl}},
\end{equation}
and the corresponding POVM elements
\begin{equation}
M_{kl} = G^{-1/2} \ket{\alpha_{kl}}\!\bra{\alpha_{kl}} G^{-1/2},
\qquad
\sum_{k,l} M_{kl}
=
\mathbbm{1}.
\label{eq:PGM}
\end{equation}
The resulting probabilities $p_{kl}=\Tr(\rho M_{kl})$ define an operational phase-space probability distribution from which the Husimi observational entropy $S_\chi^{\mathrm H}$ is computed. Its growth rate defines
\begin{equation}
\lambda_{\mathrm{OE}}^{(\mathrm{PGM})}
=
\frac{d S_\chi^{\mathrm H}}{dt}.
\label{eq:lambda_pgm}
\end{equation}

We next examine the extent to which the growth rate of observational entropy reproduces the classical Lyapunov exponent in the standard and singular kicked rotors. The quantity $\lambda_{\mathrm{OE}}^{(\mathrm{mom})}$ therefore measures the entropy production visible in momentum space alone, whereas $\lambda_{\mathrm{OE}}^{(\mathrm{PGM})}$ captures the full phase-space spreading of the quantum state at the resolution set by minimum-uncertainty coherent states. 

\begin{figure}[t]
  \centering
  \includegraphics[width=\linewidth]{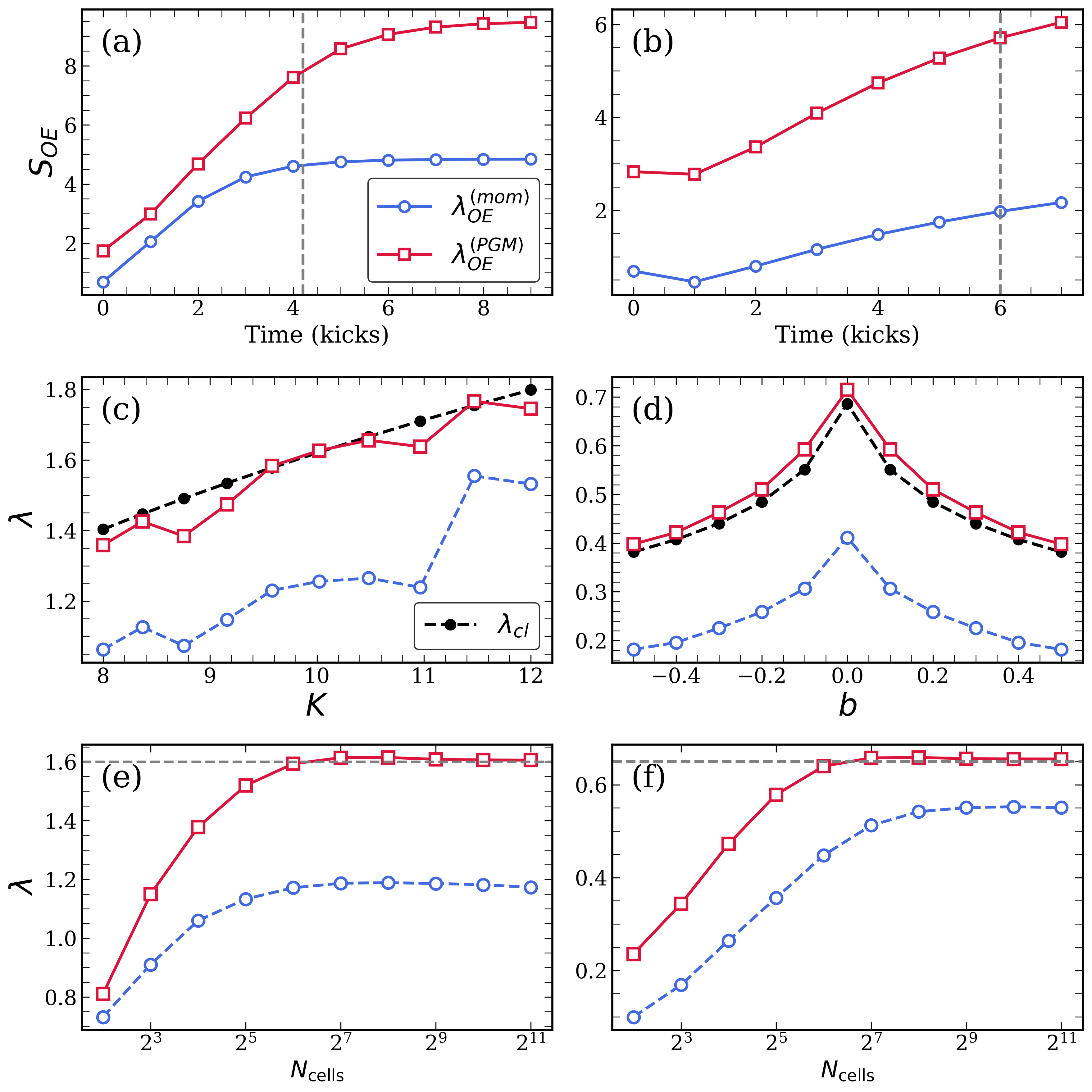}
  \caption{%
  Observational entropy growth and extracted Lyapunov exponents for the 
  Quantum Kicked Rotor (QKR, left column) and Singular Kicked Rotor (SKR, right column).
  \emph{Top row:} Time evolution of momentum-space (blue circles) and 
  PGM phase-space (red squares) observational entropy $S_{OE}$, with dashed line indicating the ehrenfest times correspondingly.
  \emph{Middle row:} Lyapunov exponents $\lambda$ as a function of kick strength $K$
  and singularity parameter $b$. Black dashed lines show the classical maximal Lyapunov exponent $\lambda_{cl}$.
  \emph{Bottom row:} Dependence of the extracted Lyapunov exponents on the number 
  of coarse-graining cells $N_{\rm cells}$ (logarithmic scale), horizontal dashed line indicating the classical lyapunov exponent.
  }
  \label{fig:oe_lyapunov}
\end{figure}

Figure~\ref{fig:oe_lyapunov} compares two coarse-graining strategies: direct binning of the momentum distribution, which yields $\lambda_{\mathrm{OE}}^{(\mathrm{mom})}$, and phase-space coarse-graining of the PGM-corrected Husimi distribution, which yields $\lambda_{\mathrm{OE}}^{(\mathrm{PGM})}$. The quantum state is initialized as a minimum-uncertainty coherent state $\ket{q_0,p_0}$ with $q_0$ sampled uniformly over the torus and $p_0=0$, and the results are averaged over 100 initial conditions. The reference classical Lyapunov exponent $\lambda_{\mathrm{cl}}$ is obtained from tangent-space evolution of the corresponding classical map.\\

The top row of Fig.~\ref{fig:oe_lyapunov} shows the time evolution of the observational entropy for the standard kicked rotor (KR) and the singular kicked rotor (SKR). In both models, the entropy exhibits a clear linear growth regime within the Ehrenfest window from which the observable Lyapunov exponent is extracted. The slope obtained from the PGM-corrected Husimi entropy, $\lambda_{\mathrm{OE}}^{(\mathrm{PGM})}$, is consistently larger than that from momentum-space coarse graining, $\lambda_{\mathrm{OE}}^{(\mathrm{mom})}$, reflecting the fact that phase-space measurements capture the full stretching of the wave packet, whereas momentum measurements probe only its projection onto a single coordinate.

The middle row compares the extracted growth rates with the classical Lyapunov exponent $\lambda_{\mathrm{cl}}$. In the QKR [Fig.~\ref{fig:oe_lyapunov}(c)], $\lambda_{\mathrm{OE}}^{(\mathrm{PGM})}$ closely follows $\lambda_{\mathrm{cl}}$ over the full range of kick strengths $K$. In the SKR [Fig.~\ref{fig:oe_lyapunov}(d)], it accurately reproduces both the magnitude and the nonmonotonic dependence of $\lambda_{\mathrm{cl}}$ on the singularity parameter $b$, including the sharp maximum near $b=0$. By contrast, $\lambda_{\mathrm{OE}}^{(\mathrm{mom})}$ captures the qualitative trends in both models but systematically underestimates the instability rate.

The bottom row shows the dependence of the extracted exponents on the number of coarse-graining cells $N_{\rm cells}$. For both models, $\lambda_{\mathrm{OE}}^{(\mathrm{PGM})}$ increases rapidly with observational resolution and saturates once $N_{\rm cells}\gtrsim 2^7$, beyond which further refinement produces negligible improvement. This saturation identifies the minimum measurement resolution required to recover the classical Lyapunov exponent quantitatively. Momentum-space estimates exhibit a similar trend but saturate at substantially smaller values. The existence of a finite resolution threshold is experimentally significant: accurate Lyapunov exponents can be obtained without arbitrarily fine phase-space reconstruction, substantially reducing the measurement resources required to quantify chaotic dynamics.

\section{Experimental Realization of Pretty Good Measurements}
\label{subsec:pgm_experiment}

A key advantage of the PGM construction is that it defines a genuine positive-operator-valued measure (POVM). Consequently, the phase-space probabilities used to compute the PGM-corrected observational entropy correspond to physically measurable outcomes rather than to a purely formal numerical construction. Since the growth rate of this entropy converges to the classical Lyapunov exponent, the PGM provides a direct experimental route to quantifying chaotic instability from finite-resolution phase-space measurements.

Starting from a set of coherent states $\{|\alpha_i\rangle\}$ with prior weights $q_i$, the frame operator is
\begin{equation}
\mathcal S
=
\sum_i q_i |\alpha_i\rangle\langle \alpha_i|.
\end{equation}
The PGM defines the POVM elements
\begin{equation}
E_i
=
\mathcal S^{-1/2}
q_i
|\alpha_i\rangle\langle \alpha_i|
\mathcal S^{-1/2},
\qquad
\sum_i E_i
=
\mathbbm{1}.
\end{equation}
For a state $\rho$, the probability of outcome $i$ is
\begin{equation}
p_i^{\mathrm{PGM}}
=
\Tr(E_i \rho).
\end{equation}
By Naimark's theorem, any POVM can be implemented as a projective measurement on an enlarged Hilbert space \cite{neumark1940spectral,peres1991optimal}. Introducing an ancilla initialized in $|0\rangle_A$, there exists a unitary $U_{SA}$ such that
\begin{equation}
U_{SA}
\bigl(
|\psi\rangle_S \otimes |0\rangle_A
\bigr)
=
\sum_i
E_i^{1/2}
|\psi\rangle_S
\otimes
|i\rangle_A .
\end{equation}
A projective measurement of the ancilla in the basis $\{|i\rangle_A\}$ then yields outcome $i$ with probability
\begin{equation}
p_i^{\mathrm{PGM}}
=
\Tr(E_i \rho).
\end{equation}

% For the KR, the outcomes are labeled by phase-space points $(q_k,p_l)$. Repeating the experiment yields the PGM-corrected Husimi probabilities $p_{kl}^{\mathrm{PGM}}(t)$, which are coarse-grained into macrostates $C_\chi$,
% \begin{equation}
% P_i(t)
% =
% \sum_{(k,l)\in C_i}
% p_{kl}^{\mathrm{PGM}}(t),
% \end{equation}
% and used to compute the observational entropy
% \begin{equation}
% S_\chi^{\mathrm{PGM}}(t)
% =
% -\sum_i
% P_i(t)
% \ln\!\left(
% \frac{P_i(t)}{V_i}
% \right).
% \end{equation}
% Within the Ehrenfest window,
% \begin{equation}
% S_\chi^{\mathrm{PGM}}(t)
% \simeq
% S_0
% +
% \lambda_{OE}^{\mathrm{PGM}} t,
% \end{equation}
% so that
% \begin{equation}
% \lambda_{OE}^{\mathrm{PGM}}
% =
% \frac{dS_\chi^{\mathrm{PGM}}}{dt}.
% \end{equation}

% In practice, one may measure coherent-state overlaps using standard Husimi tomography and apply the frame correction $\mathcal{S}^{-1/2}$ in post-processing, which yields the same probabilities as the full PGM. This makes the protocol directly compatible with existing platforms, including ultracold atoms, photonic systems, and superconducting circuits. In the semiclassical regime, we find
% \begin{equation}
% \lambda_{OE}^{\mathrm{PGM}}
% \to
% \lambda_{\mathrm{cl}},
% \qquad
% \hbar \to 0,
% \end{equation}
% establishing the PGM-based observational entropy as a direct and experimentally accessible estimator of the classical Lyapunov exponent.

\begin{figure}[t]
    \centering
    \includegraphics[width=\linewidth]{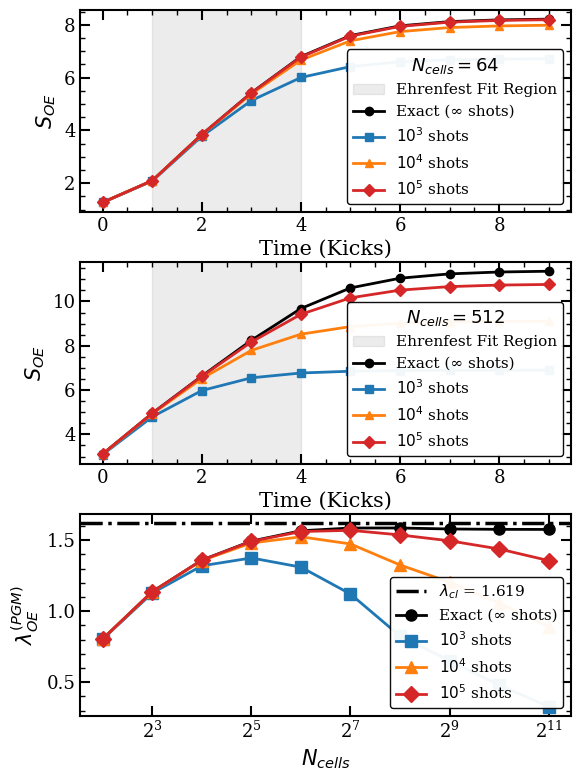}
    \caption{
    Effect of finite measurement statistics on the PGM-based observable Lyapunov exponent for the quantum kicked rotor at $K=10$. Top and middle panels: time evolution of the observational entropy for $N_{\rm cells}=2^6$ and $N_{\rm cells}=2^9$, respectively. Shaded regions indicate the Ehrenfest-time fitting window used to extract $\lambda_{OE}^{\rm (PGM)}$. Bottom panel: extracted $\lambda_{OE}^{\rm (PGM)}$ as a function of the number of phase-space cells for different numbers of measurement shots. Black circles denote the exact result and the dash-dotted line marks the classical Lyapunov exponent, $\lambda_{\rm cl}=1.619$. Increasing the number of shots extends the range of resolutions over which the PGM-based estimate converges to the classical value.
    }
    \label{fig:oe_lyapunov_shots}
\end{figure}

Figure~\ref{fig:oe_lyapunov_shots} quantifies the effect of finite measurement statistics on the extraction of the observable Lyapunov exponent from the PGM-based observational entropy. For each phase-space resolution $N_{\rm cells}$, the exact PGM probabilities are sampled from a multinomial distribution using $N_{\rm shots}$ repetitions, and the resulting entropy is fitted within the shaded Ehrenfest window to obtain $\lambda_{OE}^{\rm (PGM)}$. The top two panels show the time evolution of the observational entropy for two representative resolutions, $N_{\rm cells}=2^6$ (top) and $N_{\rm cells}=2^9$ (middle). For the coarser partition, all shot numbers closely reproduce the exact entropy growth, indicating that statistical fluctuations are negligible when many counts contribute to each cell. For the finer partition, the estimate obtained with $10^3$ shots deviates noticeably from the exact curve, while $10^4$ and $10^5$ shots recover both the linear growth regime and the correct slope. This illustrates the increasing sensitivity to shot noise as the number of phase-space cells is increased.\\

The bottom panel shows the extracted $\lambda_{OE}^{\rm (PGM)}$ as a function of $N_{\rm cells}$. For small $N_{\rm cells}$, the coarse partition under-resolves the phase-space stretching and systematically underestimates the classical Lyapunov exponent. As $N_{\rm cells}$ increases, the estimate improves and approaches the exact value. At a finite shot number, however, the estimate eventually deteriorates once the average number of counts per cell, $\bar n \sim \frac{N_{\rm shots}}{N_{\rm cells}}$, becomes too small. This leads to an optimal resolution that shifts to larger $N_{\rm cells}$ as $N_{\rm shots}$ increases. With $10^5$ shots, the extracted exponent remains close to the exact result over nearly the full range shown.\\

These results demonstrate that accurate Lyapunov exponents can be obtained from a finite number of measurements by balancing phase-space resolution against statistical uncertainty. Importantly, the optimal resolution is finite, implying that arbitrarily fine phase-space reconstruction is unnecessary for quantitative estimates of chaotic instability.

\section{Conclusions}
\label{sec:conclusions}

We have shown that observational entropy (OE) provides a unified and operational framework for extracting quantitative signatures of chaos, localization, and ergodicity-breaking transitions directly from finite-resolution measurements. Because OE depends only on coarse-grained measurement probabilities, it is experimentally accessible and does not require full state tomography, explicit eigenstate reconstruction, or nonlocal correlators.

We first demonstrated that OE can reliably capture ergodicity-breaking transitions from both long-time dynamics and eigenstate or Floquet-state ensembles. In the quantum kicked rotor, the gradual destruction of KAM tori produces the sharpest signature in the second derivative of the entropy, whereas in the Aubry--Andr\'e model the self-dual localization transition appears directly in the first derivative. The corresponding derivative extrema yield accurate estimates of the transition points over a broad range of coarse-graining scales and converge systematically toward the exact values as the system size increases. These results show that critical points can be determined reliably without requiring arbitrarily fine resolution. Once the number of coarse-graining cells exceeds a finite threshold, further refinement yields almost a little to no additional improvement, substantially reducing the measurement resources required to identify transition points.\\

We then showed that, within the Ehrenfest time regime, OE grows linearly in time and its slope defines an observable Lyapunov exponent. A phase-space coarse-graining based on the PGM-corrected Husimi distribution reproduces the classical Lyapunov exponent quantitatively in both the standard and singular kicked rotors, while momentum-space coarse graining captures only qualitative trends. Because the PGM defines a genuine POVM and can be implemented through an ancilla-assisted measurement via Naimark dilation, the corresponding phase-space observational entropy is directly accessible in experiment. Finally, we quantified the interplay between observational resolution and finite measurement statistics. Increasing the number of phase-space cells improves the estimate by resolving finer structures, but only up to an optimal resolution set by the available number of shots. Beyond this point, shot noise dominates and the accuracy deteriorates. This trade-off shows that quantitative estimates of Lyapunov exponents can be obtained with a finite and experimentally realistic number of measurements.\\

Taken together, these results establish observational entropy as a practical bridge between quantum information, statistical mechanics, and nonlinear dynamics. Within a single experimentally accessible framework, OE identifies critical points, distinguishes mechanisms of ergodicity breaking, and quantitatively recovers classical instability rates from coarse-grained measurements alone. The framework introduced here is directly applicable to current quantum simulators and opens a natural route to studying many-body localization, thermalization, Floquet prethermalization, and other emergent phenomena in interacting quantum systems \cite{sunil2026localization}.

\appendix

\begin{acknowledgments}
The authors acknowledge the use of computational resources provided by the PARAM Brahma supercomputing facility under the National Supercomputing Mission (NSM) at IISER Pune. MSS would like to acknowledge partial support from ARG-MATRICS grant ANRF/ARGM/2025/001831/TS.
\end{acknowledgments}

\appendix

\bibliography{ref}
\bibliographystyle{elsarticle-num}

\end{document}